\documentclass[prl,twocolumn,showpacs,amsmath,amssymb,superscriptaddress]{revtex4}
\usepackage{graphicx}
\usepackage{dcolumn}
\usepackage{bm}
\usepackage{epsfig}

\begin{document}
\title{Tuning the interlayer spacing in high $T_{c}$ superconductors:
penetration depth and two-dimensional superfluid density}
\author{P.J. Baker}
\author{T. Lancaster}
\author{S.J. Blundell} 
\affiliation{Oxford University Department of Physics, Clarendon Laboratory, Parks
Road, Oxford, OX1 3PU, UK}
\author{F.L. Pratt}
\affiliation{ISIS Facility, Rutherford Appleton Laboratory, Chilton, 
Oxfordshire OX11 0QX, UK}
\author{M.L. Brooks}
\affiliation{Oxford University Department of Physics, Clarendon Laboratory,  Parks
Road, Oxford, OX1 3PU, UK}
\author{S.-J. Kwon}
\affiliation{Advanced Materials Laboratory, Samsung Advanced Institute of 
Technology, Giheung-Gu, Yongin-Si, Gyeonggi-Do, 446-712, Korea}

\begin{abstract}
Substantial control of the interlayer spacing in Bi-based high temperature
superconductors has been achieved through the intercalation of guest molecules
between the superconducting layers. Measurements using implanted muons 
reveal that the penetration depth increases with increasing layer separation
while $T_{c}$ does not vary appreciably, demonstrating that the bulk
superfluid density is not the determining factor controlling $T_{c}$. 
Our results strongly suggest that the superfluid
density appearing in the Uemura scaling relation $n_{\mathrm{s}}/m^{*} \propto T_{c}$
should be interpreted as the two dimensional density within the 
superconducting layers, which we find to be constant for each class
of system investigated. 
\end{abstract}
\maketitle

\begin{figure}
\begin{center} 
\epsfig{file=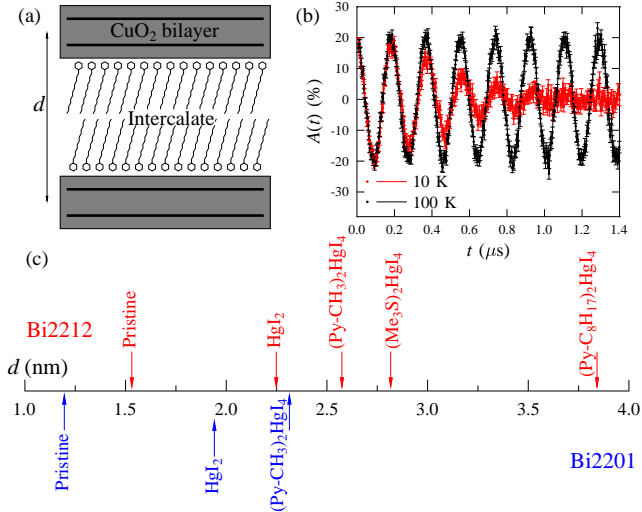,width=8.5cm}
\caption{(a) Schematic of the Bi2212 intercalates with molecular groups
acting to separated the CuO$_{2}$ bilayers. 
(b) Example TF $\mu^{+}$SR spectra measured in 40~mT above and below $T_{c}$
for polycrystalline Bi2212. The strong dephasing
for $T < T_{c}$ is caused by distribution of fields in the vortex lattice. 
(c) Interlayer/interbilayer separation $d$ for each intercalate of the
Bi2212 and Bi2201 series.
\label{data}}
\end{center}
\end{figure}

A fundamental property of high temperature superconductors (HTSs)
is the presence of CuO$_{2}$ planes  with some degree of weak 
coupling between them \cite{leggett}. The nature of this interlayer coupling \cite{history} has been 
fiercely debated in the past 
and recent advances in elucidating the Fermi surface topology of the cuprates \cite{fs1,fs2,fs3}
make an understanding of the effective dimensionality of the the HTSs a priority.
In order to address the question of dimensionality experimentally we have
modulated the interlayer coupling of the CuO$_{2}$ planes in two HTS systems by intercalating 
molecular groups between the layers. This allows us
to approach the two dimensional (2D) limit without perturbing the
superconducting oxide block. Here we present the results of a muon
spin rotation ($\mu^{+}$SR)  investigation
of the evolution of the penetration depth (and hence $n_{\mathrm{s}}/m^{*}$)
with layer separation. We are able to demonstrate that the bulk superfluid 
density $n_{\mathrm{s}}$ is not the parameter controlling 
$T_{\mathrm{c}}$ and 
that the superfluid density appearing in the Uemura relation
 $n_{\mathrm{s}}/m^{*} \propto T_{c}$ should be interpreted as the superfluid density
within the 2D superconducting layers, which is independent of the layer separation.

\begin{figure}
\begin{center}
\epsfig{file=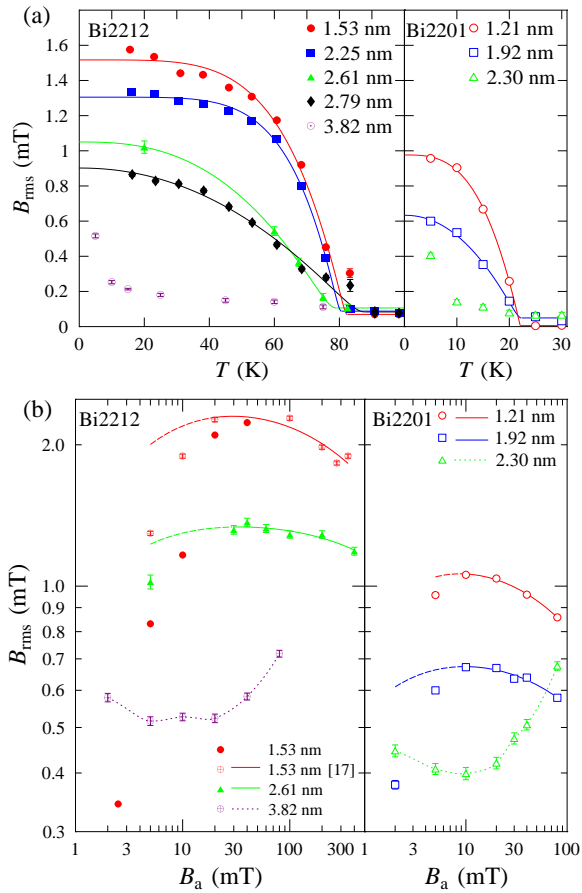,width=8.3cm}
\caption{(a) Temperature dependence of the magnetic field distribution width $B_{\mathrm{rms}}$ 
 for Bi2212 (left) and Bi2201 (right)
intercalates with $B_{\mathrm{a}}=40$~mT. Fits are described in the main text. 
(b) The evolution of $B_{\mathrm{rms}}$ with varying applied field at $T=5$~K (Bi2201) and $T=20$~K (Bi2212) 
shows
the general trend of a rounded peak in applied field. Additional data for pristine
Bi2212 were obtained from Ref.~\onlinecite{zimmermann}.
The results of the fits (described in the main text) are shown with a solid line for the range 
over which the fitting was carried out; extrapolations are shown with a dashed line.
The materials with the largest layer separations show markedly different behavior in both 
temperature and applied field (dotted lines are guides to the eye).
 \label{figa}}
\end{center}
\end{figure}

The bismuth based cuprates possess weakly bound Bi$_{2}$O$_{2}$ double layers that allow
the intercalation of guest molecules without any substantial change in the
superconducting block \cite{choy94,choy98,kwon,kwon2,kwon3,kwon4}. 
Weak interaction between the guest molecule and host layers ensures that the intercalation
process does not distort the superconducting layers. This overcomes a substantial difficulty
with alternative attempts to control the interlayer spacing \cite{fischer}, which involve 
the sequential deposition of superconducting and insulating layers.
Moreover, the intercalation of long chain organic molecules 
(shown schematically in Fig.~\ref{data}(a)) \cite{choy98,kwon} allows the possibility of
tuning the layer (or bilayer) separation $d$ on a  scale fixed by the length of the hydrocarbon chain.
Organic guest species electrically isolate the layers since hydrocarbons 
are, in general, very good insulators. 
In this paper we examine intercalation series based on
two classes of Bi-based superconductors, namely
bilayer Bi$_{2}$Sr$_{1.5}$Ca$_{1.5}$Cu$_{2}$O$_{8+\delta}$ (Bi2212) and 
single layer Bi$_{2}$Sr$_{1.6}$La$_{0.4}$CuO$_{6+\epsilon}$ (Bi2201). 
Previous work on these classes \cite{choy94,choy98,kwon,kwon2,kwon3,kwon4} 
has suggested that while the intercalation of guest species 
allows $d$ to be varied by a factor of three, the transition temperature of the
intercalated materials does not vary appreciably compared to the pristine compound.  

Transverse field muon spin rotation (TF $\mu^{+}$SR) provides a means of
accurately measuring the internal magnetic field distribution caused by
 the vortex lattice (VL) in a type II superconductor \cite{sonier}.
The vortex state of the pristine polycrystalline Bi2212 and Bi2201 systems have been
studied with $\mu^{+}$SR previously \cite{weber,russo,zimmermann}.
In a TF $\mu^{+}$SR experiment spin polarized muons are implanted in the bulk of
a superconductor, in the presence of a magnetic field $B_{c1} < B < B_{c2}$, 
which is applied perpendicular to the
initial muon spin direction. The muons stop at random positions on the length scale of the 
VL where they precess about the total local magnetic field $B$
at the muon site (mainly due to the VL), with
frequency $\omega_{\mu} = \gamma_{\mu} B$, where $\gamma_{\mu}$
is the muon gyromagnetic ratio ($= 2 \pi \times 135.5$~MHz~T$^{-1}$). 
The observed property of the experiment is the time evolution of the
muon spin polarization $P_{x}(t)$, which allows the determination of 
the distribution of local magnetic fields across the sample volume
$p(B)$, via 
\begin{equation}
P_{x}(t) = \int_{0}^{\infty} p(B) \cos (\gamma_{\mu} B t + \phi) \mathrm{d}B,
\label{pol}
\end{equation}
where $\phi$ is a phase offset associated with a particular detector
geometry.

Powder samples of several intercalates of Bi2212 and Bi2201 were prepared as
reported previously \cite{choy94,choy98,kwon,kwon2}. The intercalates studied and the separation $d$
of CuO$_{2}$ layers (or bilayers) achieved are given in Fig.~\ref{data}(c) and Table~\ref{table1}.
$\mu^{+}$SR measurements were made at the ISIS Facility using the MuSR 
spectrometer and the S$\mu$S facility using the GPS spectrometer. Samples were pressed into
pellets and mounted on a hematite backing plate to reduce the background signal.

\begin{table*}
   \caption{Properties of the intercalates of Bi$_{2}$Sr$_{1.5}$Ca$_{1.5}$Cu$_{2}$O$_{8+\delta}$ (Bi2212)
and Bi$_{2}$Sr$_{1.6}$La$_{0.4}$CuO$_{6+\epsilon}$ (Bi2201), with parameters derived from the
fitting routines described in the main text (Py=pyridine, Me=methyl). 
Fitted values are given for  Eq.~(\ref{one}) and the
fitting routine described in the main text. 
Fits were not possible for the materials with largest $d$ from each class, although the values of $T_{c}$ 
for these materials (derived from magnetization measurements) were found to 
coincide with those of the respective series to within $\sim 1$~K\cite{choy98}. 
\label{table1}}
     \begin{ruledtabular}
   \begin{tabular}{ccccccccc} 
Intercalant  & $d$~(nm) & $T_{c}$~(K)& $r$ &$B_{\mathrm{VL}}(0)$~(mT)&$B_{\mathrm{bg}}$~(mT)&$\lambda_{\mathrm{ab}}$~(nm) & $\xi$~(nm) & $d/\lambda_{\mathrm{ab}}^{2}$~($\times 10^{4}$~m$^{-1}$) \\\hline
\hline
Bi2212  & & & & & & & & \\
\hline
none                                              & 1.53  & 82.4(2) &4.6(6)&1.52(5)&0.07(2)& 200(6)  & 6(1) & 3.8(2) \\
$[$HgI$_{2}$$]$$_{0.5}$                          & 2.25 & 80.6(1) &5.7(4)&1.30(3)&0.09(1)& 230(10) & -- & 4.3(4)\\
$[$(Py-CH$_{3}$)$_{2}$HgI$_{4}$$]$$_{0.35}$      & 2.61 & 79.0(1) &2.6(1)&1.04(1)&0.11(1)& 271(7)  & 3(1) & 3.6(2)\\
$[$(Me$_{3}$S)$_{2}$HgI$_{4}$$]$$_{0.34}$        & 2.79 & 88(3)   &2.1(3)&0.90(6)&0.07(1)& 270(8) & --  & 3.8(2)\\
$[$(Py-C$_{8}$H$_{17}$)$_{2}$HgI$_{4}$$]$$_{0.35}$& 3.82 &  --     &  --  & --    & --    & --       & --  & -- \\
\hline
\hline
Bi2201 & & & & &\\
\hline
none                       & 1.21    & 22.1(2) &3.0(2)&0.98(3)&0.006(3)& 290(8)  & 13(3) &1.44(8)\\
(HgI$_{2}$)$_{0.5}$        & 1.92    & 22.3(5) &2.1(3)&0.63(5)&0.05(3) & 370(10) & 10(2) &1.40(8)\\
$[$(Py-CH$_{3}$)$_{2}$HgI$_{4}$$]$$_{0.35}$   & 2.30    &  --     &--     & --   & --     & --        & --  & --   \\
   \end{tabular}
     \end{ruledtabular}
\end{table*}

Fig.~\ref{data}(b) shows example TF $\mu^{+}$SR spectra measured above and below $T_{c}$ for 
pristine Bi2212
in an applied field of 40~mT. Above $T_{c}$ some slight broadening of the spectrum is
attributable to dephasing of the muon spins
caused by randomly directed nuclear moments near
the muon stopping sites. Below $T_{c}$ the spectrum broadens considerably due to the
dephasing contribution from the VL. 
The expected width of the field distribution in the powder sample was
estimated using the numerical results reported for the field width 
in the Ginzburg-Landau (GL) model that were obtained for a field normal to the
layers \cite{brandt}. A polycrystalline average was taken under the
assumption that the length scales $\lambda$ and $\xi$ diverge following
$1/\cos \theta$ as the field orientation approaches the plane at $\theta=0$
(high anisotropy limit), with the corresponding contribution to the
overall width scaling as $\cos \theta$.  
Fitting our data with these results allowed us to extract the second moment of the
distribution, $B_{\mathrm{rms}}$, at each measured field and temperature, and relate it to the
vortex properties and penetration depth of the material.

Fig.~\ref{figa}(a) shows the temperature dependence of $B_{\mathrm{rms}}$ in an
applied field of 5~mT for each intercalate studied. For both series, the magnitude of $B_{\mathrm{rms}}$
falls continuously with increasing temperature to a low
constant value for $T \geq T_{c}$.  The exceptions are the materials
with the largest layer separation $d$ in each series (discussed below). 
Fits are shown to an empirical power law functional form \cite{zimmermann}
\begin{equation}
B_{\mathrm{rms}}^{2}(T) = B^{2}_{\mathrm{VL}}(0) [1-(T/T_{c})^{r}]^{2} + B_{\mathrm{bg}}^{2}
\label{one}
\end{equation}
where $B_{\mathrm{VL}}(0)$ is the zero temperature contribution from the VL and 
$B_{\mathrm{bg}}$ represents a background contribution from the dipole fields
from nearby nuclei.
The parameters derived from these fits are given in table~\ref{table1}. 
There is little variation in $T_{c}$ while $B_{\mathrm{VL}}(0)$ decreases smoothly with increasing 
layer spacing. The curvature parameter $r$ falls within the range $2 \leq r \leq 6$, which is a slightly larger
range than that previously observed when this approach was applied to
 YBa$_{2}$Cu$_{3}$O$_{x}$ (YBCO) \cite{zimmermann}. The background contribution $B_{\mathrm{bg}}$
is within the range found for pristine materials
\cite{zimmermann,weber}, confirming that the intercalant is not introducing significant broadening. 

The magnetic field dependence of $B_{\mathrm{rms}}$ \cite{note1} measured at low temperature for each series
is shown in Fig.~\ref{figa}(b). 
The general trend is for $B_{\mathrm{rms}}$ to increase sharply 
with increasing field before describing a broad peak and decreasing for high fields. 
Again the exceptions to the general behavior occur for the largest layer
separation $d$.

To extract the penetration depth $\lambda_{\mathrm{ab}}$ from our measurements 
we compare the field dependence of $B_{\mathrm{rms}}$ with the behaviour expected for the
GL model (Fig.~\ref{figa}(b)) \cite{brandt}. 
The model describes the data well 
for fields around the peak value of $B_{\mathrm{rms}}$ and above, but provides a 
poor description of the data at low fields (probably due to the anisotropic magnetization and the
distribution of demagnetizing factors in the differently aligned crystallites).
We note that the $\xi$ dependence of the theoretical line width
becomes very weak near the peak, making an accurate assignment of
$\xi$ unimportant in our discussion of $\lambda_{\mathrm{ab}}$.
Indeed, close to the expected peak in $B_{\mathrm{rms}}$, for $\kappa \geq 70$
(as is the case for our systems), the GL
model predicts
the well known result $B^{2}_{\mathrm{rms}} = 0.00371 \Phi_{0}^{2}\lambda^{-4}$ \cite{brandt,sonier}.
(Here, the penetration depth $\lambda$ should be interpreted as the effective penetration depth
which includes contributions from supercurrents both parallel and perpendicular to the
layers.) The  values of $\lambda_{\mathrm{ab}}$ and $\xi$ derived from our data are given in Table~\ref{table1}\cite{note2}.
As above, we were not able to model the samples with the largest $d$ for
each class of material. 

The general trend revealed by our measurements is a clear increase in $\lambda_{\mathrm{ab}}$
with increasing layer separation $d$.
In the clean limit, we expect 
$\lambda^{-2} = \mu_{0} e^{2} n_{\mathrm{s}}/m^{*}$, where
$n_{\mathrm{s}}$ is the bulk superfluid density; thus $n_{\mathrm{s}}/m^{*}$ decreases with increasing
$d$, but crucially $T_{c}$ remains almost unchanged.
This strongly 
suggests that the bulk superfluid density is not the determining factor controlling $T_{c}$ in these systems. 
We note that our results differ from those involving varying the thickness of
a PrBa$_{2}$Cu$_{3}$O$_{7}$ (PBCO) layer in PBCO/YBCO superlattices, where it was found that 
$T_{c} \propto 1/d$ \cite{uemura97,fischer}. We stress, however, that chemical intercalation
should be expected to produce better isolated layers than those found in superlattice systems (see above). 

\begin{figure}
\begin{center}
\epsfig{file=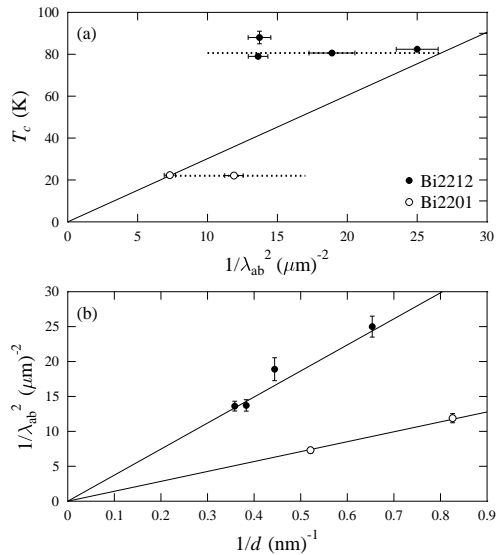,width=8cm}
\caption{
(a) Uemura plot of $T_{c}$ versus $1/\lambda_{\mathrm{ab}}^{2}$ for our compounds. The solid line represents the
scaling proposed for the underdoped cuprates \cite{uemura1}. Dotted lines are guides to the eye showing
the observed departure from the Uemura scaling.
(b) The variation of the penetration depth $\lambda_{\mathrm{ab}}$ with the inverse
layer separation $1/d$ gives a straight line dependence for each series, suggesting
the constancy of $d/\lambda_{\mathrm{ab}}^{2} (\propto  n_{\mathrm{s}}^{\mathrm{2D}}/m^{*})$.
\label{figc}}
\end{center}
\end{figure}

The behaviour we observe contrasts with that expected from a simple application of the Uemura scaling relation 
$T_{c} \propto n_{\mathrm{s}}/m^{*}$ \cite{uemura1} which holds for many underdoped 
superconductors (Fig~\ref{figc}(a)). 
This scaling relation is, however, often interpreted as suggesting the strong 
two dimensionality of the superfluid in underdoped cuprate superconductors \cite{uemura2}. 
We might, therefore, expect that for the intercalated systems 
$T_{c} \propto n_{\mathrm{s}}^{\mathrm{2D}}/m^{*} $, where $n_{\mathrm{s}}^{\mathrm{2D}}$
is the superfluid density in the superconducting layers.
If we assume that the 
superfluid density in each CuO$_{2}$ layer is the same for each class of material 
(i.e.\ a constant for Bi2212 and a different constant for Bi2201), then we should expect that 
$d/\lambda_{\mathrm{ab}}^{2} \propto n_{\mathrm{s}}^{\mathrm{2D}}/m^{*} \propto T_{c}$. 
We see from Table~\ref{table1} and Fig.~\ref{figc}(b) that our data are quite well described by 
this assumption: $d/\lambda_{\mathrm{ab}}^{2}$ is approximately constant for each class of materials, 
as is $T_{c}$. This strongly suggests that the superfluid density appearing
in the Uemura relation should be interpreted as the 2D density in the superconducting planes. 

Some insight into the meaning of our results may be obtained 
if we assume that underdoping destroys superconductivity at a 
quantum critical point (QCP) in these systems. 
Near a QCP \cite{kopp,hetel} an energy scale such as $T_{c}$ should 
vary as $T_{c} \propto \delta^{z \nu}$, 
where $\delta$ is the difference in doping from the QCP, $z$ 
is the dynamic exponent
and $\nu$ the the correlation length exponent. Combining this
with the prediction of Josephson scaling near a QCP, 
$n_{\mathrm{s}}(0) \propto \delta ^{(z+D-2) \nu}$, we
have $T_{c} \propto n_{\mathrm{s}}(0)^{z/(z+D-2)}$. For $D=2$ this predicts 
Uemura scaling $T_{c} \propto n^{\mathrm{2D}}_{s}(0)$ \cite{kopp}. 
Our findings may
therefore be suggestive of the influence of 2D quantum fluctuations in
the underdoped regime. 

Finally we note that VL breakup effects explain the anomalous
behavior of the samples with the largest layer separation (see
Fig.~2). A sharp drop in $B_{\rm rms}(T)$, similar to that
observed in our Bi2212 ($d=3.82$~nm) and Bi2201
($d=2.3$~nm) samples, has previously been observed in pristine
Bi2212 \cite{lee97}, where it was attributed to VL melting at
small superfluid densities close to $T_c$. Both electromagnetic
and Josephson coupling are expected to stabilize the VL in these
materials \cite{blat,lee97}. The temperature at which thermal fluctuations
break up a VL stabilized solely by electromagnetic coupling is
given by $T_{em} = \Phi^{2}_{0} \bar{c}/(32 k_{\rm B} \mu_0 \pi^{2}
\lambda_{\mathrm{ab}}^{2})$ \cite{clem91} (where $\bar{c}$ is the average CuO$_{2}$ layer
separation), which is constant for each series as $T \rightarrow 0$.
We can therefore estimate $T_{\mathrm{em}} \sim 14$~K for Bi2212 and
$T_{\mathrm{em}} \sim 10$~K for Bi2201, both in reasonable agreement
with the temperatures at which we observe a sharp drop in
$B_{\rm rms}(T)$ in the samples with the largest layer spacing. This
suggests that the Josephson coupling, which decreases strongly
with increasing layer spacing, stabilizes the VL to higher
temperature in the samples with smaller layer spacing but is
too weak to do so in the two samples with the largest $d$. This
leaves only electromagnetic coupling and the observed VL breakup
for $T>T_{\mathrm{em}}$. 

 Part of this work was carried out at the S$\mu$S, Paul Scherrer
Institut, CH and at the ISIS facility, Rutherford Appleton Laboratory, UK. 
We thank E.H. Brandt for access to the results of his calculations for the GL model.
We are grateful to Alex Amato for technical assistance and Ted Forgan for
useful discussions.
This work is supported by the EPSRC (UK) and by the European Commission
under the Sixth Framework Programme through the Key Action:
Strengthening the European Research Area, Research Infrastructures,
Contract No.\ RII3-CT-2003-505925.
T.L. acknowledges support from the Royal Commission for the Exhibition
of 1851. S. J. Kwon is very grateful for the support by the Korea-UK Scientific 
Networking Program (M6-0405-00-0001).


\begin{thebibliography}{xx}
\bibitem{leggett}
A.J. Leggett, Nature Phys. {\bf 2}, 134 (2006).

\bibitem{history}
J.M. Wheatley, T.C. Hsu and P.W. Anderson, Science {\bf 333}, 121 (1988);
S. Doniach and W.E. Lawrence, Proc 12th Int. Conf. Low Temp. Phys.
(ed E. Kanda) 361 (Keigaku, Tokyo, 1971).

\bibitem{fs1}
N.E. Hussey {\it et al.}, Nature {\bf 425}, 814 (2003).

\bibitem{fs2}
N. Doiron-Leyraud {\it et al.}, Nature {\bf 447}, 565 (2007).

\bibitem{fs3}
N. Harrison, R.D. McDonald and J. Singleton, 
Phys. Rev. Lett. {\bf 99}, 206406 (2007).

\bibitem{choy94}
J.-H. Choy {\it et al.}, J. Am. Chem. Soc. {\bf 116}, 11564 (1994).

\bibitem{choy98}
J.-H. Choy, S.J. Kwon and K.S. Park, Science, {\bf 280}, 1589 (1998).

\bibitem{kwon}
S.-J. Kwon and D.Y. Jung, Solid State Commn. {\bf 130}, 287 (2004).

\bibitem{kwon2}
S.-J. Kwon {\it et al.}, Phys. Rev. B {\bf 66}, 224510 (2002).


\bibitem{kwon3}
S.-J. Kwon  and J.-H. Choy, Inor. Chem. {\bf 42}, 8134 (2003).

\bibitem{kwon4}
S.-J. Kwon {\it et al.}, Supercond. Sci. Technol. {\bf 18}, 470 (2005).

\bibitem{fischer}
\O. Fischer {\it et al.}, Physica B {\bf 169}, 116 (1991).

\bibitem{sonier}
J.E. Sonier, J.H. Brewer and R.F. Kiefl, Rev. Mod. Phys. {\bf 72}, 769 (2000).

\bibitem{weber}
M. Weber {\it et al.}, Phys. Rev. B {\bf 48}, 13022 (1993).

\bibitem{russo}
P.L. Russo {\it et al.}, Phys. Rev. B {\bf 75}, 054511 (2007).


\bibitem{zimmermann}
P. Zimmermann {\it et al.}, Phys. Rev. B {\bf 52}, 541 (1995).

\bibitem{brandt}
E.H. Brandt, Phys. Rev. B {\bf 68}, 054506 (2003).


\bibitem{note1}
For the $d=2.25$~nm and 2.79~nm  materials, which were measured at ISIS, 
the resolution limit that results from the 
ISIS pulse width prevents data being measured at fields $> 100$~mT.

\bibitem{note2}
For intercalated Bi2212 with $d=2.25$~nm  and 2.79~nm we may estimate
$\lambda_{\mathrm{ab}}$ from the points near the largest measured $B_{\mathrm{rms}}$. 
This is reasonable since these materials have layer separations
between the extreme values of $d$ for our samples, so we would expect to be able to interpolate their behaviour.


\bibitem{uemura97}
Y.J. Uemura, Physica C {\bf 282-287}, 194 (1997).

\bibitem{uemura1}
Y.J. Uemura {\it et al.}, Phys. Rev. Lett. {\bf 62}, 2317 (1989). 

\bibitem{uemura2}
Y.J. Uemura {\it et al.}, Phys. Rev. Lett. {\bf 66}, 2665 (1991).


\bibitem{kopp}
A. Kopp and S. Chakravarty, Nature Phys. {\bf 1}, 54 (2005).

\bibitem{hetel}
I. Hetel, T.R. Lemberger and M. Randeria, Nature Phys. {\bf 3}, 700 (2007).


\bibitem{lee93}
S.L. Lee {\it et al.},  Phys. Rev. Lett. {\bf 71}, 3862 (1993).

\bibitem{blat}
G. Blatter {\it et al}, Phys. Rev. B {\bf 54}, 72 (1996).

\bibitem{lee97}
S.L. Lee {\it et al.}, Phys. Rev B, {\bf 55}, 5666 (1997).

\bibitem{clem91}
J.R. Clem, Phys. Rev. B {\bf 43}, 7837 (1991).

\end{thebibliography}
\end{document}